\documentclass[a4paper,prd,twocolumn,aps,preprintnumbers,showpacs]{revtex4}
\usepackage{amssymb}
\usepackage{epsfig}
\usepackage{graphicx}
\usepackage{dcolumn}
\usepackage{bm}
\usepackage{hyperref}
 
\def \be  {\begin{equation}}
\def \ee  {\end{equation}}
\def \beq  {\begin{equation}}
\def \eeq {\end{equation}}
\def \ba  {\begin{eqnarray}}
\def \ea  {\end{eqnarray}}
\def \baa {\begin{eqnarray*}}
\def \eaa {\end{eqnarray*}}
\def \bb  {\begin{thebibliography}}
\def \eb  {\end{thebibliography}}

\def \nn {\nonumber}

\def \lab #1 {\label{#1}}

\newcommand\as{\alpha_s}

\def \fracs #1#2 {\mbox{\small $\frac{#1}{#2}$}}

\def \bin #1#2 {{\left({#1}\atop{#2}\right)}}
\def\lapproxeq{{\ \lower 0.6ex \hbox{$\buildrel<\over\sim$}\ }}
\def\gapproxeq{{\ \lower 0.6ex \hbox{$\buildrel>\over\sim$}\ }}

\def\hepph  #1 {{hep-ph/#1 }}

\begin{document}
%

\title{Threshold resummation for polarized (semi-)inclusive deep inelastic scattering}
\author{Daniele P. Anderle, Felix Ringer, Werner Vogelsang}
\affiliation{Institute for Theoretical Physics, 
T\"ubingen University, Auf der Morgenstelle 14, 72076 T\"ubingen, Germany}
\begin{abstract}
We explore the effects of  the resummation of large logarithmic perturbative corrections 
to double-longitudinal spin asymmetries for inclusive and semi-inclusive deep inelastic 
scattering in fixed-target experiments. We find that the asymmetries are overall rather robust 
with respect to the inclusion of the resummed higher-order terms. Significant effects are observed
at fairly high values of $x$, where resummation tends to decrease the spin asymmetries. 
This effect turns out to be more pronounced for semi-inclusive scattering. We also 
investigate the potential impact of resummation on the extraction of polarized valence
quark distributions in dedicated high-$x$ experiments.
\end{abstract}

\date{\today}
\maketitle

\section{Introduction \label{intro}}

Longitudinal double-spin asymmetries in inclusive and semi-inclusive deep inelastic scattering 
have been prime sources of information on the nucleon's spin structure for several decades.
They may be used to extract the helicity parton distributions of the nucleon, 
\ba
\label{pPDF}
\Delta f(x,Q^2)\equiv f^+(x,Q^2)-f^-(x,Q^2)\ ,
\ea
where $f^+$ and $f^-$ are the distributions of parton $f=q,\bar{q},g$ with 
positive and negative helicity, respectively, when the parent nucleon has positive
helicity. $x$ denotes the momentum fraction of the parton and $Q$ the hard scale at
which the distribution is probed. Inclusive polarized deep inelastic scattering (DIS), $\vec{\ell}\vec{p}\rightarrow \ell X$,
offers access to the combined quark and antiquark distributions for a given flavor,
$\Delta q+\Delta \bar{q}$, whereas in semi-inclusive deep inelastic scattering (SIDIS), 
$\vec{\ell}\vec{p}\rightarrow \ell h X$, one exploits the fact that a produced hadron $h$ (like 
a $\pi^+$) may for instance have a quark of a certain flavor as a valence quark,  but not the corresponding 
antiquark~\cite{Frankfurt:1989wq}. In this way, it becomes possible to separate quark and antiquark distributions
in the nucleon from one another, as well as to better determine the distributions for the various flavors. 
HERMES~\cite{Airapetian:2004zf} and recent COMPASS~\cite{SIDIShelicityCompass2010} 
measurements have marked significant progress concerning the accuracy and kinematic coverage of 
polarized SIDIS measurements. The inclusive measurements have improved vastly as 
well~\cite{hermesinkl,DIShelicityCompass2010,Zheng:2003un,clas,Burkardt:2008jw}. 
Some modern analyses of spin-dependent
parton distributions include both inclusive and semi-inclusive 
data~\cite{deFlorianpPDF,deFlorian:2011cr,Leader:2010rb}. In addition, high-precision data for polarized 
SIDIS will become available from experiments to be carried out at the Jefferson Lab after the CEBAF upgrade to 
a 12 GeV beam~\cite{Burkert:2012rh}. Here the focus will be on the large-$x$ regime. 

A good understanding of the theoretical framework for the description of spin asymmetries in lepton
scattering is vital for a reliable extraction of polarized parton distributions. 
In a recent paper~\cite{AnderleFelixWerner} we have investigated the effects of QCD threshold
resummation on hadron multiplicities in SIDIS in the HERMES and COMPASS kinematic regimes.
SIDIS is characterized by two scaling variables, Bjorken-$x$ and a variable $z$ given by the energy 
of the produced hadron over the energy of the virtual photon in the target rest frame. Large logarithmic 
corrections to the SIDIS cross section arise when the corresponding partonic variables become
large, corresponding to scattering near a phase space boundary, where real-gluon emission is suppressed.
This is typically the case for the 
presently relevant fixed-target kinematics. Threshold resummation addresses these logarithms to all
orders in the strong coupling. In~\cite{AnderleFelixWerner} we found fairly significant resummation 
effects on the spin-averaged multiplicities. Since the spin-dependent cross section is subject to similar logarithmic 
corrections as the unpolarized one, 
it is worthwhile to explore the effects of resummation on the spin asymmetries. This is the goal of the present paper. 
Our calculations will be carried out both for inclusive DIS and for SIDIS. We note that previous 
work~\cite{Simula:2001iy,Osipenko:2005nx} has addressed the large-$x$ resummation for the inclusive 
spin-dependent structure function $g_1$, with a focus on the moments of $g_1$ and their $Q^2$-dependence. 
In this paper we are primarily concerned with spin asymmetries and with semi-inclusive scattering.

Our work will use the framework developed in~\cite{AnderleFelixWerner}. In Section \ref{sec:A1}, we briefly review the basic 
terms and definitions relevant for longitudinal spin asymmetries, and we describe the extension of threshold resummation to the 
polarized case. In Section~\ref{sec:results} our phenomenological results are presented. We compare our
resummed inclusive and semi-inclusive spin asymmetries with available HERMES, COMPASS and Jefferson Lab data. We also
discuss the relevance of resummation for the extraction of $\Delta u/u$ and $\Delta d/d$ at large values of $x$. 

\section{Resummation for Longitudinal Spin Asymmetries in DIS and SIDIS \label{sec:A1}}
 
\subsection{Leading and next-to-leading order expressions \label{ssec:basic}}
 
We first consider the polarized SIDIS process $\vec{\ell}(k)\vec{p}(P)\to \ell(k')h(P_h)X$ with longitudinally  polarized beam and target 
and with an unpolarized hadron in the final state. The corresponding double-spin asymmetry is given by a ratio of
structure functions~\cite{Airapetian:2004zf}:
\be
\label{A1h}
A^h_1(x,z,Q^2)\approx\frac{g^h_1(x,z,Q^2)}{F_1^h(x,z,Q^2)}\;,
\ee
where $Q^2=-q^2$ with $q$ the momentum of the virtual photon, 
$x=Q^2/(2P\cdot q)$ is the usual Bjorken variable, and $z\equiv P\cdot P_h/P\cdot q$ the corresponding 
hadronic scaling variable associated with the fragmentation process. 

Using factorization, the polarized structure function $g_1^h$, which appears in the numerator of Eq.~(\ref{A1h}), 
can be written as
\begin{eqnarray}
\label{g1hallorders}
2g_1^h(x,z,Q^2) &=&\sum_{f,f'=q,\bar{q},g} 
\int_x^1 \frac{d\hat{x}}{\hat{x}}\int_z^1 \frac{d\hat{z}}{\hat{z}}\, \Delta f \left(\frac{x}{\hat{x}},
\mu^2\right) \,\nn\\[2mm]
&\times&D^h_{f'} \left(\frac{z}{\hat{z}},\mu^2\right)\,\Delta{\cal{C}}_{f'f}
\left(\hat{x},\hat{z},\frac{Q^2}{\mu^2},\alpha_s(\mu^2)\right) ,\nn \\
\end{eqnarray}
where $\Delta f(\xi,\mu^2)$ denotes the polarized distribution function for parton $f$ of Eq.~(\ref{pPDF}), 
whereas $D^h_{f'} \left(\zeta,\mu^2\right)$ 
is the corresponding fragmentation function for parton $f'$ going to the observed hadron $h$. The $\Delta{\cal{C}}_{f'f}$ 
are spin-dependent coefficient functions. We have set all factorization and renormalization scales equal and collectively denoted 
them by $\mu$. In~(\ref{g1hallorders}) $\hat{x}$ and $\hat{z}$  are the partonic counterparts of the hadronic variables $x$ and $z$. 
Setting for simplicity $\mu=Q$, we use the short-hand-notation
\be\label{g1hallorders1}
2g_1^h(x,z,Q^2) \,\equiv\,\sum_{f,f'=q,\bar{q},g} \left[\Delta f\otimes \Delta{\cal{C}}_{f'f}\otimes D^h_{f'}\right](x,z,Q^2)
\ee
for the convolutions in~(\ref{g1hallorders}). A corresponding expression for the ``transverse'' unpolarized structure 
function $2F_1^h$ can be written by replacing the polarized parton distributions with the unpolarized ones, and using
unpolarized coefficient functions which we denote here by ${\cal{C}}_{f'f}$.

The spin-dependent hard-scattering coefficient functions $\Delta{\cal{C}}_{f'f}$ in~(\ref{g1hallorders}) 
can be computed in perturbation theory:
\beq\label{Cpert}
\Delta{\cal{C}}_{f'f}\,=\,\Delta C^{(0)}_{f'f}+\frac{\alpha_s(\mu^2)}{2\pi}\Delta C^{(1)}_{f'f}+{\cal O}(\alpha_s^2)\;.
\eeq
At leading order (LO), we have
\be
\Delta{\cal{C}}_{qq}(\hat{x},\hat{z})\,=\,\Delta{\cal{C}}_{\bar{q}\bar{q}}
(\hat{x},\hat{z})\,=\, e_q^2\,\delta(1-\hat{x})\delta(1-\hat{z})\,,
\ee
with the quark's fractional charge $e_q$. All other coefficient functions vanish. 
The same result holds for the LO coefficient function for the spin-averaged structure function $2F_1^h$. 
Hence the asymmetry in Eq.~(\ref{A1h}) reduces to 
\be
\label{paronmodelA1h}
A^h_1=\frac{\sum\limits_{q} e_q^2\left[\Delta q(x,Q^2)D^h_q(z,Q^2)+\Delta\bar{q}(x,Q^2)D^h_{\bar{q}}
(z,Q^2)\right]}{\sum\limits_{q} e_q^2\left[q(x,Q^2)D^h_q(z,Q^2)+\bar{q}(x,Q^2)D^h_{\bar{q}}
(z,Q^2)\right]}\; .
\ee
At next-to-leading order (NLO), Eq.~(\ref{g1hallorders}) becomes
\ba
\label{g1h}
2g_1^h(x,z,Q^2)&=&\nn\\[2mm]
&&\hspace*{-2.6cm}\sum_q e^2_q \bigg\{\Delta q(x,Q^2)D^h_q(z,Q^2)+\bar{q}(x,Q^2)D^h_{\bar{q}}(z,Q^2)\nn\\[2mm]
&&\hspace*{-2.6cm}+\frac{\as(Q^2)}{2\pi}\left[\left(\Delta q\otimes D^h_q+
\Delta \bar{q}\otimes D^h_{\bar{q}}\right)\otimes \Delta C^{(1)}_{qq}\right. \nn\\[2mm]
&&\hspace*{-1.25cm}+\left(\Delta q+\Delta \bar{q}\right)\otimes \Delta C^{(1)}_{gq}\otimes D^h_g\nn\\[2mm]
&&\hspace*{-1.3cm}\left. +\Delta g\otimes \Delta C^{(1)}_{qg}\otimes (D_q^h+D_{\bar{q}}^h)\right](x,z,Q^2)\bigg\},
\ea
where the symbol $\otimes$ denotes the convolution defined in Eqs.~(\ref{g1hallorders}),(\ref{g1hallorders1}). 
The explicit expressions for
the spin-dependent NLO coefficients $\Delta C^{(1)}_{f'f}$ have been derived in~\cite{deFlorian:1997zj,roth}. 
The corresponding spin-averaged NLO coefficient functions $C^{(1)}_{f'f}$ may be found
in~\cite{altarelli,Nason:1993xx,fupe,graudenz,AnderleFelixWerner,deFlorian:1997zj,roth}.

In the case of {\it inclusive} polarized DIS, the longitudinal spin asymmetry $A_1$ is given in analogy with~(\ref{A1h}) 
by
\be
\label{A1}
A_1(x,Q^2)\approx\frac{g_1(x,Q^2)}{F_1(x,Q^2)}\;.
\ee
The inclusive structure functions $g_1$ and $F_1$ have expressions analogous to their SIDIS counterparts,
except for the fact that they do not contain any fragmentation functions, of course. The unpolarized and polarized NLO 
coefficient functions for inclusive DIS may be found at many places; see, for example~\cite{fupe,Gluck:1995yr}.

\subsection{Threshold resummation \label{resA1}}

As was discussed in~\cite{AnderleFelixWerner}, the higher-order terms in the spin-averaged
SIDIS coefficient function ${\cal{C}}_{qq}$
introduce large terms near the ``partonic threshold'' $\hat{x}\to 1$,  $\hat{z}\to 1$. The same is true for the 
spin-dependent $\Delta{\cal{C}}_{qq}$. At NLO, choosing again for simplicity the scale $\mu=Q$, one has
\begin{eqnarray}
\label{sidiseq8res}
\nonumber
\Delta C_{qq}^{(1)}(\hat{x},\hat{z}) &\sim& e_q^2 C_F
\Bigg[ \\
\nonumber\\ \nonumber
&&\hspace*{-2cm}+\,2 \delta(1-\hat{x}) \left(\frac{\ln(1-\hat{z})}{1-\hat{z}}\right)_+ 
+2 \delta(1-\hat{z}) \left(\frac{\ln(1-\hat{x})}{1-\hat{x}}\right)_+  \\[2mm]
&&\hspace*{-2cm}+\,\frac{2}{(1-\hat{x})_+(1-\hat{z})_+}-8\delta(1-\hat{x})\delta(1-\hat{z})\Bigg],
\end{eqnarray}
where the ``+''-distribution is defined as usual. The expression on the right-hand side is in fact
identical to the one for the unpolarized coefficient function near threshold~\cite{AnderleFelixWerner}.
At the $k$th order of perturbation theory, the coefficient function contains terms of the form 
$\alpha_s^k\delta(1-\hat{x}) \left(\frac{\ln^{2k-1}(1-\hat{z})}{1-\hat{z}}\right)_+$,
$\alpha_s^k\delta(1-\hat{z}) \left(\frac{\ln^{2k-1}(1-\hat{x})}{1-\hat{x}}\right)_+$,
or ``mixed'' distributions $\alpha_s^k \left(\frac{\ln^{m}(1-\hat{x})}{1-\hat{x}}\right)_+
\left(\frac{\ln^{n}(1-\hat{z})}{1-\hat{z}}\right)_+$ with $m+n=2k-2$, plus
terms less singular by one or more logarithms. Again, each of these terms will appear
equally in the unpolarized and in the polarized coefficient function. The reason for this is that 
the terms are associated with emission of soft gluons~\cite{AnderleFelixWerner}, which does
not care about spin. Threshold resummation addresses the large logarithmic terms to all orders
in the strong coupling. The resummation for the case of SIDIS was carried out in~\cite{AnderleFelixWerner}.
Given these results and the equality of the spin-averaged and spin-dependent coefficient
functions near threshold, it is relatively straightforward to perform the resummation for the 
polarized case. Having the resummation for both $g_1^h$ and $F_1^h$, we obtain
resummed predictions for the experimentally relevant spin asymmetry $A_1^h$.

In~\cite{AnderleFelixWerner,Sterman:2006hu,Cacciari:2001cw} threshold resummation for SIDIS was derived using an 
eikonal approach, for which exponentiation of the threshold logarithms 
is achieved in Mellin space. One takes Mellin moments of $g_1^h$ separately in the two independent 
variables $x$ and $z$~\cite{altarelli,Stratmann:2001pb}:
\begin{equation}
\label{momresug1h}
\tilde{g}^h_1(N,M,Q^2)\equiv\int_0^1 dx x^{N-1}\int_0^1 dz z^{M-1}\,
g^h_1(x,z,Q^2).
\end{equation}
With this definition, Eq.~(\ref{g1hallorders1}) takes the form (again at scale $\mu=Q$)
\ba
\label{resug1h}
2\tilde{g}_1^h(N,M,Q^2)&=&
\sum_{f,f'=q,\bar{q},g} \Delta \tilde{f}^N(Q^2)  \nn\\[2mm]
&\times&\Delta\tilde{\cal{C}}_{f'f}(N,M,\alpha_s(Q^2))\tilde{D}^{h,M}_{f'}(Q^2)\,,\nn \\
\ea
where the moments of the polarized parton distributions and the fragmentation functions are defined as
\ba
\Delta \tilde{f}^N(Q^2)&\equiv &\int_0^1 dx x^{N-1}\Delta f(x,Q^2),\nn\\[2mm]
\tilde{D}_{f'}^{h,M}(Q^2)&\equiv &\int_0^1 dz z^{M-1}D^h_{f'}(z,Q^2),
\ea
and the double Mellin moments of the polarized coefficient functions are
\ba
\Delta\tilde{{\cal C}}_{f'f}\left(N,M,\alpha_s(Q^2)\right)&\equiv&
\int_0^1 d\hat{x}\hat{x}^{N-1}\int_0^1 d\hat{z}\hat{z}^{M-1}\,\nn\\[2mm]
&\times& \Delta{\cal{C}}_{f'f}
\left(\hat{x},\hat{z},1,\alpha_s(Q^2)\right) \,.\nn\\
\ea
Large $\hat{x}$ and $\hat{z}$ in $\Delta{\cal{C}}_{f'f}$ correspond to large 
$N$ and $M$ in $\Delta\tilde{{\cal C}}_{f'f}$, respectively. 

The resummed spin-dependent coefficient function is identical to the
spin-averaged one of~\cite{AnderleFelixWerner} and reads to next-to-leading
logarithmic (NLL) accuracy in the $\overline{{\mathrm{MS}}}$-scheme:
\ba\label{resummed4}
&&\hspace*{-11mm}\Delta\tilde{\cal{C}}^{{\mathrm{res}}}_{qq}(N,M,\alpha_s(Q^2))=e_q^2 
H_{qq}\left(\alpha_s(Q^2)\right) \nn\\[2mm]
&&\hspace*{-8mm}\times\exp\left[2 \int_{\frac{Q^2}{\bar{N}\bar{M}}}^{Q^2} 
{dk_\perp^2\over k_\perp^2}A_q\left(\as(k_\perp^2)\right)
 \ln \left(\frac{k_{\perp}}{Q}\sqrt{\bar{N} \bar{M}}\right) \right],
\ea
where $\bar{N}\equiv N{\mathrm{e}}^{\gamma_E}$, $\bar{M}\equiv 
M{\mathrm{e}}^{\gamma_E}$, with $\gamma_E$ the Euler constant, and
\begin{equation}\label{exp_A}
 A_q(\alpha_s) = \frac{\alpha_s}{\pi} A_q^{(1)} + 
\left(\frac{\alpha_s}{\pi}\right)^2 A_q^{(2)}+ \dots
\end{equation}
is a perturbative function. The coefficients required to NLL read
\begin{equation}
 A_q^{(1)} = C_F, \quad A_q^{(2)} = \frac{1}{2} C_F\left[C_A
\left(\frac{67}{18}-\frac{\pi^2}{6}\right) - \frac{5}{9} N_f\right],
\end{equation}
where $C_F=4/3$, $C_A=3$ and $N_f$ is the number of active flavors.
Furthermore,
\be
H_{qq}\left(\alpha_s\right)=1+\frac{\alpha_s}{2\pi}C_F
\left( -8 + \frac{\pi^2}{3}\right) +{\cal O}(\alpha_s^2)\;.
\ee
The explicit NLL expansion of the exponent in~(\ref{resummed4}) is
given by~\cite{AnderleFelixWerner}
\ba\label{h1h2a}
&&\hspace*{-2mm}\int_{\frac{Q^2}{\bar{N}\bar{M}}}^{Q^2} 
{dk_\perp^2\over k_\perp^2}A_q\left(\as(k_\perp^2)\right)
 \ln \left(\frac{k_{\perp}}{Q}\sqrt{\bar{N} \bar{M}}\right)\nn\\[2mm]
 &&\approx
h_q^{(1)}\left(\frac{\lambda_{NM}}{2}\right)\,\frac{\lambda_{NM}}{2b_0\alpha_s(\mu^2)}
+h_q^{(2)}\left(\frac{\lambda_{NM}}{2}, \frac{Q^2}{\mu^2},\frac{Q^2}{\mu_F^2}
\right),\nn\\[2mm]
\ea
where
\ba\label{h1h2}
\lambda_{NM} &\equiv&b_0\alpha_s(\mu^2)\left(\log\bar{N}+\log\bar{M}\right),\nn\\[2mm]
h_q^{(1)}(\lambda) &=& \frac{A_q^{(1)}}{2 \pi b_0 \lambda}\left[2\lambda + 
(1- 2\lambda) \ln(1-2\lambda)\right],\nonumber \\[2mm]
h_q^{(2)}\left(\lambda, \frac{Q^2}{\mu^2},\frac{Q^2}{\mu_F^2}\right) &= & -
\frac{A_q^{(2)}}{2 \pi^2 b_0^2} \left[2\lambda + \ln (1-2\lambda)\right] \nn\\[2mm]
&&\hspace*{-2cm} +\frac{A_q^{(1)}b_1}{2\pi b_0^3}\left[2\lambda + 
\ln(1-2\lambda)+ \frac{1}{2} \ln^2(1-2\lambda)\right] \nonumber  \\[2mm]
& &\hspace*{-2cm} + \frac{A_q^{(1)}}{2\pi b_0} \left[2 \lambda + 
\ln(1-2\lambda)\right] \ln\frac{Q^2}{\mu^2}  - \frac{A_q^{(1)}}{\pi b_0} \lambda
 \ln \frac{Q^2}{\mu_F^2},\nn\\[2mm]
 \ea
with
\ba
b_0 &=& \frac{11 C_A - 4 T_R N_f}{12\pi},\nn\\[2mm]
b_1 &=&\frac{17 C_A^2 - 10 C_A T_R N_f -6 C_F T_R N_f}{24\pi^2}.
\ea
The functions $h_q^{(1)}$, $h_q^{(2)}$ collect all leading-logarithmic
and NLL terms in the exponent, which are of the form 
$\alpha_s^k \ln^n \bar{N} \ln^m \bar{M}$ with 
$n+m=k+1$ and $n+m=k$, respectively. Note that we have restored the full dependence
on the factorization and renormalization scales in the above expressions. 

The polarized moment-space structure function $\tilde{g}_1^{h,\mathrm{res}}$ resummed to NLL is obtained by 
inserting the resummed coefficient function into in Eq.~(\ref{resug1h}). To get the physical hadronic structure function 
$g_1^{h,\mathrm{res}}$ one needs to take the Mellin inverse of the moment-space expression. As in~\cite{AnderleFelixWerner},
we choose the required integration contours in complex $N,M$-space according to the \textit{minimal prescription} of~\cite{Catani:1996yz},
in order to properly deal with the singularities arising from the Landau pole due to the divergence of the perturbative 
running strong coupling constant $\as$
at scale $\Lambda_{{\mathrm{QCD}}}$. Moreover, we match the resummed $g^{h,\mathrm{res}}_1$ to its NLO value, \textit{i.e.} 
we subtract the $\mathcal{O}(\alpha_s)$ expansion from the resummed expression and add the full NLO result:
\ba 
\label{match}
g^{h,\mathrm{match}}_1\equiv g^{h,\mathrm{res}}_1-g^{h,\mathrm{res}}_1\Big |_{\mathcal{O}(\as)}+
g^{h,\mathrm{NLO}}_1\;.
\ea
The final resummed and matched expression for the spin asymmetry $A_1^h$ is then given by
\begin{equation}
\label{resA1h}
A^{h,\mathrm{res}}_1(x,z,Q^2)\equiv\frac{g^{h,\mathrm{match}}_1(x,z,Q^2)}{F_{1}^{h,\mathrm{match}}(x,z,Q^2)}\;.
\end{equation}

Similar considerations can be made for inclusive DIS, where again the resummation for $g_1$ proceeds identically to that
of $F_1$ in moment space. Only single Mellin moments of the structure function have to be taken:
\begin{equation}
\label{momresug1}
\tilde{g}_1(N,Q^2)\equiv\int_0^1 dx x^{N-1}\,
g_1(x,Q^2).
\end{equation}
The threshold resummed coefficient function is the same as in the spin-averaged case and is discussed for example
in~\cite{AnderleFelixWerner}. We note that the outgoing quark in the process $\gamma^* q\to q$ remains 
``unobserved'' in inclusive DIS. At higher orders this is known to generate Sudakov {\it suppression} 
effects~\cite{cmn} that counteract the Sudakov enhancement associated with soft-gluon radiation from the 
initial quark. This is in contrast to SIDIS, where the outgoing quark fragments and hence is ``observed'',
so that both the initial and the final quark contribute to Sudakov enhancement. As a result, resummation effects
are generally larger in SIDIS than in DIS, for given kinematics.

\section{Phenomenological results \label{sec:results}}

We now analyze numerically the impact of threshold resummation on the semi-inclusive and inclusive DIS asymmetries 
$A_1^h$ and $A_1$. Given that the resummed exponents are identical for the spin-averaged and spin-dependent 
structure functions, we expect the resummation effects to be generally very modest. On the other hand, it is also clear that 
the effects will not cancel identically in the spin asymmetries: Even though the resummed exponents for $g_1$ and 
$F_1$ are identical in Mellin-moment space, they are convoluted with different parton distributions and hence
no longer give identical results after Mellin inversion. Moreover, the matching procedure also introduces differences
since the NLO coefficient functions are somewhat different for $g_1$ and $F_1$. It is therefore still relevant to investigate
the impact of resummation on the spin asymmetries. We will compare our results to data sets from HERMES~\cite{Airapetian:2004zf} 
and COMPASS~\cite{SIDIShelicityCompass2010,DIShelicityCompass2010}. In addition, we present some results relevant for 
measurements at the Jefferson Laboratory~\cite{Zheng:2003un,clas}, 
in particular those to be carried out in the near future after the CEBAF upgrade to 
12 GeV~\cite{Burkert:2012rh}. 
\begin{figure}
\vspace*{-2mm}
\hspace*{-0.8cm}
\includegraphics[width=0.55\textwidth,angle=90]{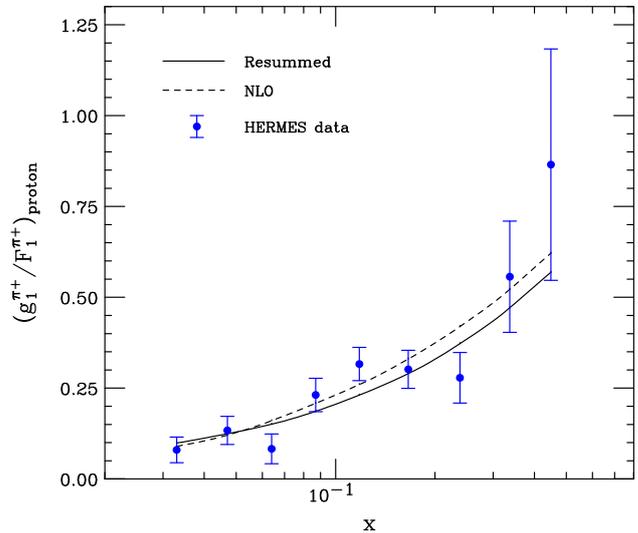}
\vspace*{-1.5cm}
\caption{\label{fig1}\sf Spin asymmetry for semi-inclusive $\pi^+$ production off a proton target. The data points are
from~\cite{Airapetian:2004zf} and show statistical errors only. The $\langle x \rangle$ and $\langle Q^2 \rangle$ values 
were taken accordingly to the HERMES measurements.} 
\end{figure}
\begin{figure}
\vspace*{-6mm}
\hspace*{-0.8cm}
\includegraphics[width=0.55\textwidth,angle=90]{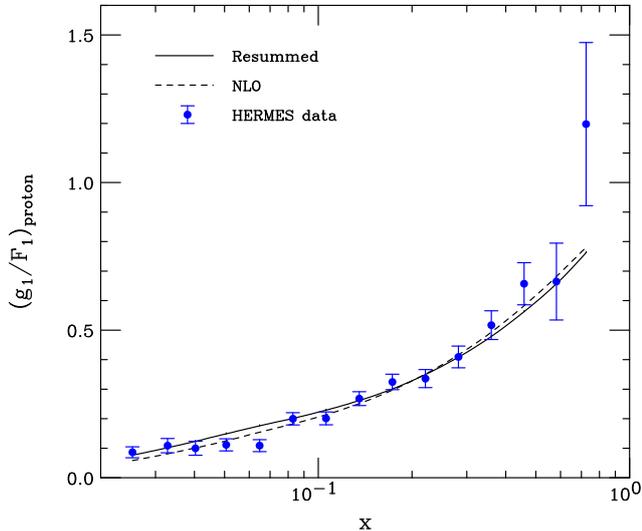}
\vspace*{-1.5cm}
\caption{\label{fig2}\sf  Spin asymmetry for inclusive polarized DIS off a proton target. The data points are
from~\cite{hermesinkl} and show statistical errors only. The $\langle x \rangle$ and $\langle Q^2 \rangle$ values 
were taken accordingly to the HERMES measurements.}
\end{figure}

For our calculations we use the NLO polarized parton distribution functions 
of~\cite{deFlorianpPDF} and the unpolarized ones of~\cite{Martin:2002aw}. Our choice of the latter is motivated
by the fact that this set was also adopted as the baseline unpolarized set in~\cite{deFlorianpPDF}, so that the
two sets are consistent in the sense that the same strong coupling constant is used. Additionally, in the case of SIDIS 
we choose the ``de Florian-Sassot-Stratmann''~\cite{defloriandss} NLO set of fragmentation functions. In this work, 
we choose to focus only on pions in the final state. Resummation effects for other hadrons will be very similar.
The factorization and renormalization scales are set to $Q$. 

Figures~\ref{fig1} and~\ref{fig2} present comparisons of our resummed calculations with HERMES 
data~\cite{Airapetian:2004zf} for semi-inclusive ($\pi^+$) and inclusive DIS, respectively, both off a proton target
at $\sqrt{s}\approx{7.25}$ GeV.
The error bars show the statistical uncertainties only. For the SIDIS asymmetry, we integrate the numerator and 
the denominator of Eq.~(\ref{A1h}) separately over a region of $0.2<z<0.8$. We plot the theoretical results
at the average values of $x$ and $Q^2$ of each data point and connect the points by a line. The figures
show the NLO (dashed lines) and the resummed-matched (solid lines) results. As one
can see, the higher-order effects generated by resummation are indeed fairly small, although not negligible.
They are overall more significant for SIDIS, which is expected due to the additional threshold logarithms
in SIDIS (see discussion at the end of Sec.~\ref{resA1}). We expect the resummed results to be most reliable at 
rather high values of $x\gtrsim 0.2$ or so~\cite{AnderleFelixWerner}. In this regime, there is a clear
pattern that resummation tends to decrease the spin asymmetries compared to NLO, more pronounced so for SIDIS. 
In other words, higher-order corrections enhance the spin-averaged cross section somewhat more strongly than the
polarized one.

\begin{figure}[b]
\vspace*{-2mm}
\hspace*{-0.8cm}
\includegraphics[width=0.55\textwidth,angle=90]{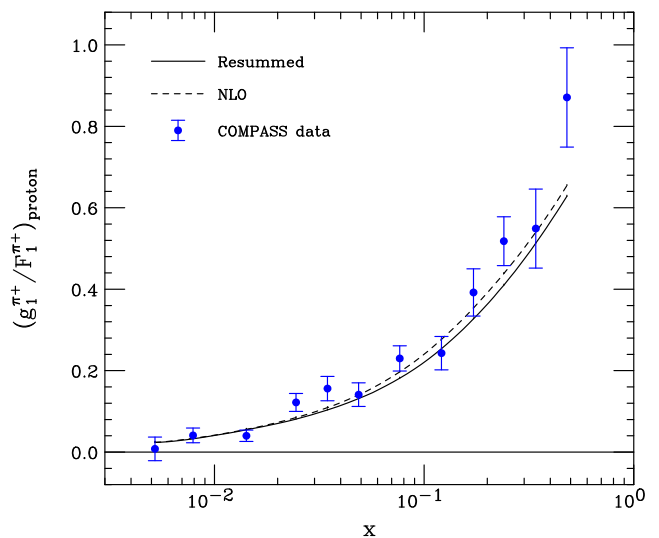}
\vspace*{-1.5cm}
\caption{\label{fig3}\sf Same as Fig.~\ref{fig1} but comparing to the COMPASS measurements~\cite{SIDIShelicityCompass2010}.}
\end{figure}

\begin{figure}
\vspace*{-2mm}
\hspace*{-0.8cm}
\includegraphics[width=0.55\textwidth,angle=90]{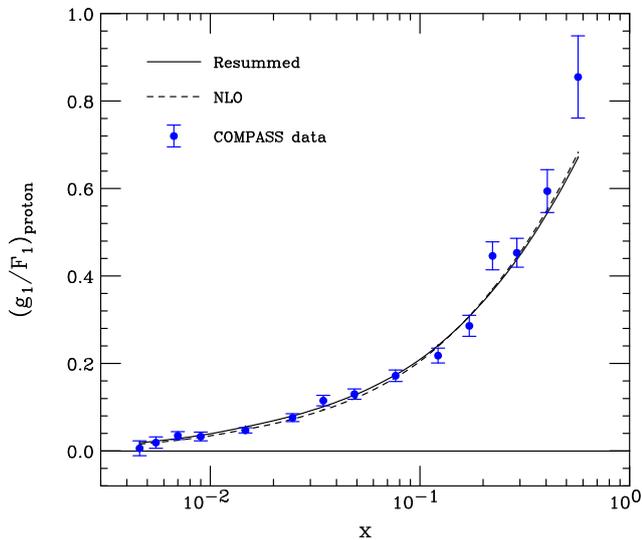}
\vspace*{-1.5cm}
\caption{\label{fig4}\sf Same as Fig.~\ref{fig2} but comparing to the COMPASS measurements~\cite{DIShelicityCompass2010}.}
\end{figure}

Figures~\ref{fig3} and~\ref{fig4} show similar comparisons to the SIDIS and DIS asymmetries measured
by COMPASS~\cite{SIDIShelicityCompass2010,DIShelicityCompass2010} with a polarized muon beam at
 $\sqrt{s}\approx$17.4 GeV. For COMPASS kinematics the effects of threshold resummation are overall 
 somewhat smaller due to the fact that one is further away from partonic threshold because of the higher center-of-mass
 energy. However, the results  remain qualitatively similar to what we observed for HERMES kinematics. 

\begin{figure}
\vspace*{-2mm}
\hspace*{-0.8cm}
\includegraphics[width=0.55\textwidth,angle=90]{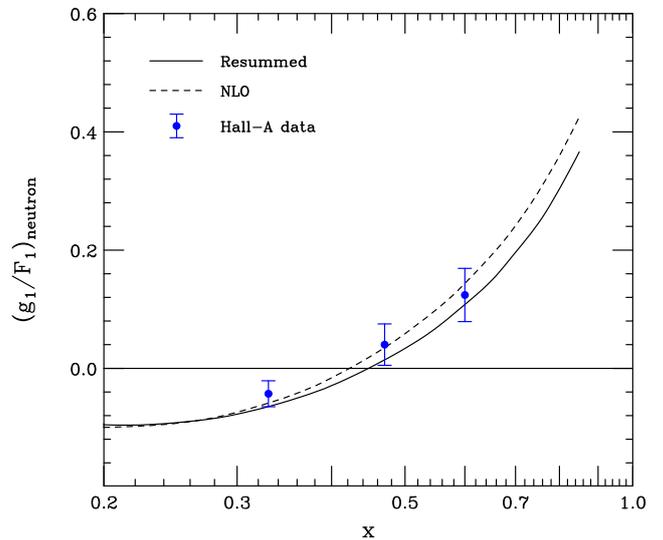}
\vspace*{-1.5cm}
\caption{\label{fig5}\sf  Spin asymmetry for inclusive polarized DIS off a neutron target. The data points are
from~\cite{Zheng:2003un} and show statistical errors only. The $Q^2$ values in the theoretical calculation
were chosen as $Q^2=x \times 8$~GeV$^2$.}
\end{figure}

The inclusive {\it neutron} spin asymmetry is particularly interesting from the point of view
of resummation, since it is known~\cite{Zheng:2003un} to exhibit a sign change at fairly
large values of $x$. Near a zero of the polarized cross section resummation effects are expected
to be particularly relevant. Figure~\ref{fig5} shows the asymmetry at NLO and for the NLL
resummed case. For illustration we show the presently most precise data available, which are
from the Hall-A Collaboration~\cite{Zheng:2003un}  at the Jefferson Laboratory. In order
to mimic the correlation of $x$ and $Q^2$ for the present Jefferson Lab kinematics,
we choose $Q^2=x \times 8$~GeV$^2$ in the theoretical calculation. As one can see,
the effects of resummation are indeed more pronounced than for the inclusive proton 
structure functions considered in Figs.~\ref{fig2} and~\ref{fig4}. Evidently the zero 
of the asymmetry shifts slightly due to resummation. On the other hand, the asymmetry
is overall still quite stable with respect to the resummed higher order corrections. 

\begin{figure}
\vspace*{-2mm}
\hspace*{-0.8cm}
\includegraphics[width=0.55\textwidth,angle=90]{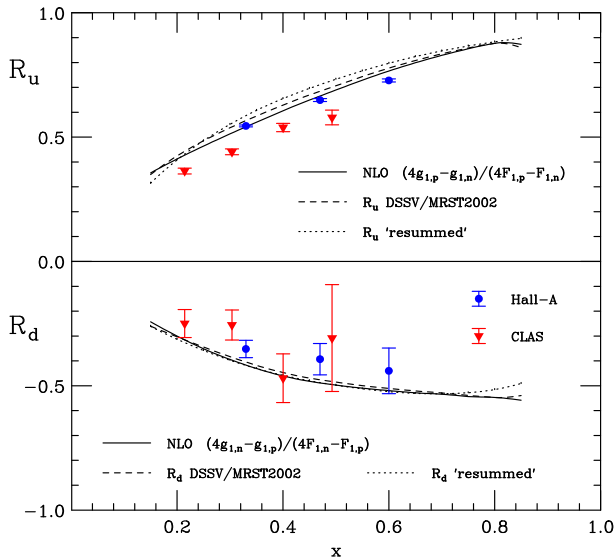}
\vspace*{-1.5cm}
\caption{\label{fig6}\sf High-$x$ up and down polarizations $(\Delta u+\Delta\bar{u})/(u+\bar{u})$ and 
$(\Delta d+\Delta\bar{d})/(d+\bar{d})$. The solid lines show the ratios of structure functions on the right-hand sides
of Eq.~(\ref{udratios}), while the dashed lines show the actual parton distribution ratios as represented by 
the NLO sets of~\cite{deFlorianpPDF} and~\cite{Martin:2002aw}. The dotted lines show the expected shift of 
the distributions when resummation effects are included in their extraction, using Eq.~(\ref{resumpdf}). 
The $Q^2$ values in the theoretical calculation
were chosen as $Q^2=x \times 8$~GeV$^2$. We also 
show the present Hall-A~\cite{Zheng:2003un} and CLAS~\cite{clas} data obtained from inclusive
DIS measurements. Their error bars are statistical only.}
\end{figure}

The latter observation is quite relevant for the extraction of polarized large-$x$ parton distributions
from data for proton and neutron spin asymmetries in lepton scattering. For instance, to good
approximation~\cite{Zheng:2003un} one may use the inclusive structure functions to directly
determine the combinations $(\Delta u+\Delta\bar{u})/(u+\bar{u})$ and 
$(\Delta d+\Delta\bar{d})/(d+\bar{d})$. At lowest order, and neglecting the
contributions from strange and heavier quarks and antiquarks, one has 
\ba\label{udratios}
R_u&\equiv&\frac{\Delta u+\Delta\bar{u}}{u+\bar{u}}(x,Q^2)\,=\,\frac{4g_{1,{\mathrm{p}}}-g_{1,{\mathrm{n}}}}
{4F_{1,{\mathrm{p}}}-F_{1,{\mathrm{n}}}}(x,Q^2)\,,\nn\\[2mm]
R_d&\equiv&\frac{\Delta d+\Delta\bar{d}}{d+\bar{d}}(x,Q^2)\,=\,\frac{4g_{1,{\mathrm{n}}}-g_{1,{\mathrm{p}}}}
{4F_{1,{\mathrm{n}}}-F_{1,{\mathrm{p}}}}(x,Q^2)\,,
\ea
where the subscripts p,n denote a proton or neutron target, respectively. One may therefore 
determine $(\Delta u+\Delta\bar{u})/(u+\bar{u})$ and $(\Delta d+\Delta\bar{d})/(d+\bar{d})$ 
directly from experiment by using measured structure functions $g_{1,{\mathrm{p}}},
g_{1,{\mathrm{n}}},F_{1,{\mathrm{p}}},F_{1,{\mathrm{n}}}$ in~(\ref{udratios}). Up to 
certain refinements required by the fact that measurements of the ratios 
$g_{1,{\mathrm{p}}}/F_{1,{\mathrm{p}}}$ and $g_{1,{\mathrm{n}}}/F_{1,{\mathrm{n}}}$ 
are more readily available than those of the individual structure functions, 
this is essentially the approach used by the Hall-A Collaboration
(alternatively, one may also use the corresponding spin asymmetry for the deuteron
instead of the neutron one~\cite{clas}).
In the following we explore the typical size of the corrections to the ratios due to higher
orders. Figure~\ref{fig6} shows first of all the structure function ratios on the right-hand side
of~(\ref{udratios}), computed at NLO using as before the polarized and unpolarized 
parton distribution functions of~\cite{deFlorianpPDF} and~\cite{Martin:2002aw}, 
respectively (solid lines). We have again chosen $Q^2=x \times 8$~GeV$^2$.
Using~(\ref{udratios}), these ratios would correspond to the ``direct experimental 
determinations'' of $R_u$ and $R_d$. The dashed lines in the figure show the
actual ratios $(\Delta u+\Delta\bar{u})/(u+\bar{u})$ and $(\Delta d+\Delta\bar{d})/(d+\bar{d})$
as given by the sets of parton distribution functions that we use. Any difference
between the solid and dashed lines is, therefore, a measure of the significance of
effects related to strange quarks and antiquarks, and to NLO corrections. As one can
see, these have relatively modest size. Finally, we estimate the potential effect of resummation 
on $R_u,R_d$: Following~\cite{sv,corc}, we define `resummed' quark (and antiquark) distributions
by demanding that their contributions to the structure functions $g_1$, $F_1$ match those of 
the corresponding NLO distributions, which is ensured by setting
\be \label{resumpdf}
\tilde{q}^{N,{\mathrm{res}}}(Q^2)\equiv\frac{\tilde{\cal{C}}^{{\mathrm{NLO}}}_q(N,\alpha_s(Q^2))}
{\tilde{\cal{C}}^{{\mathrm{res}}}_q(N,\alpha_s(Q^2))}\tilde{q}^{N,{\mathrm{NLO}}}(Q^2)
\ee
in Mellin-moment space. Here, $\tilde{\cal{C}}^{{\mathrm{NLO}}}_q$ and $\tilde{\cal{C}}^{{\mathrm{res}}}_q$
are the NLO and resummed quark coefficient functions for the inclusive structure function $F_1$, respectively. We
match the resummed coefficient function to the NLO one by subtracting out its NLO contribution and adding
the full NLO one, in analogy with~(\ref{match}). Equation~(\ref{resumpdf}) can be straightforwardly 
extended to the spin-dependent case. The ratios $R_u,R_d$ for these `resummed' parton distributions
are shown by the dotted lines in Fig.~\ref{fig6}. As one can see, they are quite close to the other
results, indicating  that resummation is not likely to induce very large changes in the parton polarizations
extracted from future high-precision data. For illustration, we also show the Hall-A~\cite{Zheng:2003un} 
and CLAS~\cite{clas} data in the figure, which have been obtained using parton-model relations for
the inclusive structure functions, similar to~(\ref{udratios}). One can see that the error bars of the data are
presently still larger than the differences between our various theoretical results. This situation is expected
to be improved with the advent of the Jefferson Lab 12-GeV upgrade~\cite{Burkert:2012rh} or an 
Electron Ion Collider~\cite{elke}. As is well-known, 
SIDIS measurements provide additional information on $R_u,R_d$, albeit so far primarily at lower 
$x$~\cite{Airapetian:2004zf}.

\section{Conclusions}

We have investigated the size of threshold resummation effects on double-longitudinal spin 
asymmetries for inclusive and semi-inclusive deep inelastic scattering in fixed-target experiments. 
Overall, the asymmetries are rather stable with respect to resummation, in particular for the
inclusive case. Towards large values of $x$, resummation tends to cause a decrease of the 
spin asymmetries, which is more pronounced in the semi-inclusive case and for 
asymmetries measured off neutron targets. 

The relative robustness of the spin asymmetries bodes well for the extraction of high-$x$ 
parton polarizations $(\Delta u+\Delta\bar{u})/(u+\bar{u})$ and $(\Delta d+\Delta\bar{d})/(d+\bar{d})$, 
which are consequently also rather robust. Nevertheless, knowledge of the predicted higher-order 
corrections should be quite relevant when future high-statistics large-$x$ data become available. 
On the theoretical side, it will be interesting to study the interplay of our perturbative corrections
with power corrections that are ultimately also expected to become important at 
high-$x$~\cite{Simula:2001iy,Osipenko:2005nx,Leader:2006xc,Blumlein:2010rn,acc}, although it
appears likely that present data are in a window where the perturbative corrections clearly
dominate. Finally, we note that related large-$x$ logarithmic effects have also been investigated
for the nucleon's light cone wave function~\cite{Avakian:2007xa}, where they turn out 
to enhance components of the wave function with non-zero orbital angular momentum,
impacting the large-$x$ behavior of parton distributions. It will be very worthwhile to 
explore the possible connections between the logarithmic corrections discussed here
and in~\cite{Avakian:2007xa}.\\

\section{Acknowledgments}
   
We thank Marco Stratmann for useful communications.   
This work was supported in part by the German Bundesministerium f\"{u}r Bildung und Forschung (BMBF),
grant no. 05P12VTCTG.

\end{document}